\documentstyle[12pt]{article}
\begin{document}
\title{ON THE LIFETIME OF A COLD DARK MATTER PARTICLE AND
 THE COSMOLOGICAL DIFFUSE PHOTON BACKGROUND}
\author{D. PALLE \\
Zavod za teorijsku fiziku, Institut Rugjer Bo\v {s}kovi\'{c} \\
P.O.Box 1016 Zagreb, Croatia}  
\date{ }
\maketitle
 We show that a Majorana heavy neutrino with a mass ${\cal O}$(100 TeV) 
 is a good candidate particle for cold dark matter. It can be 
 responsible for the majority of the cosmological diffuse photon
 background owing to lifetime of the order of ${\cal O}(10^{25}s)$,
 dominantly fixed by the radiative two-body decay.
 The lifetime is suppressed by two mechanisms: the leptonic GIM cancellation
 and the see-saw weak coupling suppression. 
 As a fermion cold dark matter particle, 
 a heavy neutrino favours the average
 mass density of the Universe constrained by the Einstein-Cartan 
 cosmology. \\
\\
\vspace{10 mm}

\section{Introduction and motivation}

During the past decades we have been witnessed
 of  great progress in 
cosmology which places  severe constraints on the physics of
 fundamental constituents of matter. There is a common belief
that it is not possible to resolve the problems of structure
formation of the Universe, basic cosmological parameters, 
nucleosynthesis or a diffuse photon background without the
introduction of cold and hot dark matter and the violation of
the baryon number.

On the other hand, measurements of the LSND and the SuperKamiokande have
definitely confirmed the existence of massive neutrinos and
their flavour mixing, urging our necessity to build a more 
predictable theory in particle physics than the Standard Model.

A diffuse photon background as a source of ionization at galactic or
extragalactic scales represents a challenge for particle physics
to search for a process that can explain the phenomenon \cite{Sciama}.
It has been known
 for a long time that the measurement of the flux of
decay-produced photons can constrain the lifetime of the decayed
particle \cite{Kolb,Moha,Sciama}.
From a detailed study of ionization fluxes \cite{Sciama},
Sciama has estimated that the
 mass and the lifetime of a decaying light neutrino
are $m_{\nu _{\tau}}\simeq 28 eV,\ \tau\simeq 10^{24}s$.
However, it is very difficult to reconcile such a high mass with the present
cosmological fits of the structure formation which require an order of
magnitude smaller neutrino masses (as hot dark matter) and cold dark
matter (that should dominate the mass density of the Universe).
The observed neutrino oscillations at the SuperKamiokande and the limits
from other detectors prohibit large masses if we exclude mass degeneracy.
The standard weak interaction calculation of the neutrino lifetime\cite{Moha} 
with a mass $m_{\nu}\simeq 30 eV$ gives $\tau\simeq 10^{36}s$, thus twelve
orders of magnitude larger than it is required by the ionization flux.
One can improve the result adding a contribution of a new scalar or
vector charged particle exchanged in the quantum loop, but new particles
could completely spoil the structure of electroweak interactions
that are verified to very high precision at LEP and SLD.
To measure a photon flux of the decaying neutrino near the Sun, an extreme
ultraviolet detector on the satellite has been proposed\cite{Sciama}.
In 1997 the satellite was launched and now there are data that 
are incompatible with Sciama's decaying neutrino\cite{satelit}.

It seems that light neutrinos are not very promising candidates to 
solve the cosmological problem of diffuse photon background.

In this paper we investigate abundances and lifetimes of heavy
neutrinos, within an electroweak theory proposed a few years ago,
in order to find a cold dark matter(CDM) particle
 whose radiative decay could be
responsible for a diffuse photon background.
 
\section{Cosmic abundances of heavy neutrinos}

To investigate the cosmological significance of any particle, one
should know its interactions and cross sections to solve 
the Boltzmann equations in curved spacetime.
Let us recall some results on the freeze-out and
abundances of heavy neutrinos.

Heavy neutrinos with masses ${\cal O}(1GeV)$ can play a role of
the CDM particle owing to high abundances calculated from the
Z-boson mediated annihilation cross section \cite{Kolb} with
the following scaling $ \sigma_{Z} \propto m_{N}^{2} $.
If neutrinos have masses much larger than weak bosons, the
authors of Ref.\cite{Enqvist} showed that the total annihilation
cross section was not dominated by the fermion pair production
$ \sigma (\bar{N}N \rightarrow \bar{f}f) \propto m_{N}^{-2} $
, but by the W-boson pair production
$ \sigma (\bar{N}N \rightarrow W^{-}W^{+}) \propto m_{N}^{2}.$
They concluded \cite{Enqvist} that "there is no cosmological
upper bound on the masses of very heavy stable neutrinos".
They assumed the standard weak coupling of
(Dirac) neutrinos. 

However, Griest and Kamionkowski \cite{Griest} successfully put
an upper bound on the mass of the CDM particle 
estimating the upper bound of the total annihilation cross section
without reference to any particular interaction model or theory.

Nevertheless, it should be remembered that heavy neutrinos
have a substantially different scaling of the cross sections in
the fermion or boson pair production. If one excludes
the $W^{-}W^{+}$ production, one can reach a cosmologically
acceptable abundance of a heavy neutrino as a dark matter particle
with a mass $m_{N}\simeq 900 GeV$ 
 \cite{Enqvist}.

It is now necessary to define our framework for a study of
heavy neutrinos. We work within an electroweak theory
proposed some years ago \cite{Palle1}. We called it (BY) theory in
that paper. It differs from the Standard Model(SM) (called
(AX) theory) in two essential ingredientes: (1) Nambu-Goldstone
scalars carry nonvanishing lepton numbers and only Majorana
fields acquire masses at tree level, (2) there is no Higgs scalar and
the principle of the noncontractible space (the existence of finite 
 UV scale) is introduced into 
the local gauge field theory as a symmetry-breaking mechanism, 
  fixing the tree-level weak boson and Majorana fermion masses.
The theory contains three light and three heavy Majorana neutrinos.
Further insights and discussions could be found in 
Ref. \cite{Palle1}. Here we concentrate on the leptonic sector.

Let us recall the structure of the four-component Dirac
spinor:

\begin{equation}
\Psi_{D}\equiv\left(\begin{array}{c}\eta_{a} \\
\dot{\xi}^{\dot{a}}\end{array}\right), 
\end{equation}
\begin{eqnarray*}
\eta_{a}\ transforms\ under\ a\ matrix\ S\ of\ SL(2,C),
\end{eqnarray*}
\begin{eqnarray*}
\dot{\xi}^{\dot{a}}\ transforms\ under\ a\ matrix\ (S^{-1})^{*}\ of\ SL(2,C).
\end{eqnarray*}

To retain all degrees of freedom,
one can choose the following set of fermion fields: 

\begin{eqnarray}
e_{L},\ e_{R},\ \Psi_{L},\ (\Psi^{C})_{L}.
\end{eqnarray}

These fields interact with $SU(2)\times U(1)$ gauge fields and
Nambu-Goldstone scalars as follows:

\begin{equation}
\begin{array}{c}
{\cal L}\ =\ {\cal L}_{lep}+{\cal L}_{g.bos}+{\cal L}_{scal}+
{\cal L}^{M}_{Yuk}+{\cal L}_{g.fix}+{\cal L}_{FP},
\end{array}
\end{equation}
$
\begin{array}{l}
{\cal L}_{lep}=\bar{R}i\gamma^{\mu}(\partial_{\mu}+ig'B_{\mu})R+
\bar{L}i\gamma^{\mu}(\partial_{\mu}+\frac{i}{2}g'B_{\mu}-
ig\frac{\tau^{i}}{2}A^{i}_{\mu})L+\bar{\psi}_{R}i\gamma^{\mu}\partial_{\mu}
\psi_{R}, \nonumber\\
{\cal L}_{g.bos}=-\frac{1}{4}F^{i}_{\mu\nu}F^{i\mu\nu}-\frac{1}{4}
B_{\mu\nu}B^{\mu\nu}, \nonumber\\
{\cal L}_{scal}=(\partial_{\mu}\Phi^{\dagger}+i\frac{g'}{2}B_{\mu}\Phi
^{\dagger}+i\frac{g}{2}\tau^{i}A^{i}_{\mu}\Phi^{\dagger})(\partial^{\mu}
\Phi-i\frac{g'}{2}B^{\mu}\Phi-i\frac{g}{2}\tau^{i}A^{i\mu}\Phi)
, \nonumber\\
{\cal L}_{Yuk}^{M}=-Y^{\psi}_{M}\bar{L^{C}}\Phi\psi_{R}+h.c.,
\nonumber\\
definitions:\ e=charged\ lepton,\ \psi=neutral\ Dirac\ lepton,\nonumber\\ 
L=\left(\begin{array}{c}\psi_{L}\\e_{L}\end{array}\right),\ R=e_{R};\nonumber\\
F^{i}_{\mu\nu}=\partial_{\mu}A^{i}_{\nu}-\partial_{\nu}A^{i}_{\mu}+
g\epsilon^{ijk}A^{j}_{\mu}A^{k}_{\nu},\ B_{\mu\nu}=\partial_{\mu}B_{\nu}-
\partial_{\nu}B_{\mu};\nonumber\\
 \Phi=\left(\begin{array}{c}\phi^{+}\\{\phi^{0}}
\end{array}\right),\ L^{C}=\left(\begin{array}{c}
(e^{C})_{L}\\(\Psi^{C})_{L}\end{array}\right),\nonumber\\
\phi^{0}(x)\ =\ (v+i\chi^{0}(x))/{\sqrt{2}},\ v=symmetry\ breaking\ 
parameter,\nonumber\\ \phi^{\pm},\chi^{0}=Nambu-
Goldstone\ scalars.\nonumber
\end{array}
$

The physical spectrum of neutral leptons looks like \cite{Palle1,Kayser}

\begin{eqnarray*}
\left( \begin{array}{cc}m_{L} & m_{D} \\ m_{D} & 0 \end{array}
\right)  \left( \begin{array}{c} a \\ b \end{array}\right) =
\lambda \left( \begin{array}{c} a \\ b \end{array}\right) , \nonumber
\end{eqnarray*}

\begin{eqnarray*}
a^{2}+b^{2}=1, 
\end{eqnarray*}

\begin{eqnarray}
\Psi_{N}=a_{N}f+b_{N}F,\ \Psi_{\nu '}=a_{\nu '}f+b_{\nu '}F, 
\end{eqnarray}
\begin{eqnarray*}
f\equiv \frac{1}{\sqrt{2}}(\Psi_{L}+(\Psi_{L})^{C}),\ 
F\equiv \frac{1}{\sqrt{2}}(\Psi_{R}+(\Psi_{R})^{C}), 
\end{eqnarray*}
\begin{eqnarray}
\lambda_{N}=\frac{1}{2}(m_{L}+\sqrt{m_{L}^{2}+4m_{D}^{2}}),\ 
\lambda_{\nu '}=\frac{1}{2}(m_{L}-\sqrt{m_{L}^{2}+
4m_{D}^{2}}), 
\end{eqnarray}
\begin{eqnarray*}
a_{N}=\frac{m_{L}+\sqrt{m_{L}^{2}+4m_{D}^{2}}}
{(8m_{D}^{2}+2m_{L}^{2}+2m_{L}\sqrt{m_{L}^{2}+4m_{D}^{2}})^   
{\frac{1}{2}}},\ b_{N}=\frac{2m_{D}a_{N}}{m_{L}+
\sqrt{m_{L}^{2}+4m_{D}^{2}}}, 
\end{eqnarray*}
\begin{eqnarray*}
a_{\nu '}=\frac{\sqrt{m_{L}^{2}+4m_{D}^{2}}-m_{L}}
{(8m_{D}^{2}+2m_{L}^{2}-2m_{L}\sqrt{m_{L}^{2}+4m_{D}^{2}})^
{\frac{1}{2}}},\ b_{\nu '}=\frac{2m_{D}a_{\nu '}}
{m_{L}-\sqrt{m_{L}^{2}+4m_{D}^{2}}}, 
\end{eqnarray*}
\begin{eqnarray*}
m_{L} \gg m_{D} \Rightarrow \lambda_{N}\simeq m_{L},\ 
\lambda_{\nu '}\simeq -\frac{m_{D}^{2}}{m_{L}},
\end{eqnarray*}
\begin{eqnarray*}
a_{N}\simeq 1,\ b_{N}\simeq \frac{m_{D}}{m_{L}},\ 
a_{\nu '}\simeq \frac{m_{D}}{m_{L}},\ 
b_{\nu '}\simeq -1, 
\end{eqnarray*}
\begin{eqnarray}
\Psi_{\nu}\equiv \gamma_{5}\Psi_{\nu '} \Rightarrow
\lambda_{\nu}=-\lambda_{\nu '}.
\end{eqnarray}

Heavy Majorana particles acquire
masses predominantly from the quantum loops with a Nambu-Goldstone   
scalar in the strong coupling regime of Dyson-Schwinger
equations \cite{Palle1}.
However, the effective strong coupling for fermions with masses
of a few TeV is saturated in bootstrap equations and its value
(like the QCD strong coupling in the infrared domain) is not so
high.
The cross section for heavy neutrinos annihilating
 into a heavy neutrino pair, calculated in the 't Hooft-Feynman
gauge is

\begin{eqnarray}
\sigma_{\chi^{0}}(N_{i}N_{i}\rightarrow N_{j}N_{j})&=&
\frac{\pi}{\sin ^{4}\theta_{W}}(\tilde{\alpha_{e}})_{i}
(\tilde{\alpha_{e}})_{j} \\
&\times &  
\frac{1}{(s-m_{\chi^{0}}^{2})^{2}+\Gamma^{2}_{\chi^{0}}
m_{\chi^{0}}^{2}}
\frac{s-2m_{N_{j}}^{2}}{s}\sqrt{(s-4m_{N_{i}}^{2})
(s-4m_{N_{j}}^{2})}. \nonumber
\end{eqnarray}

Notice that we cannot decouple Nambu-Goldstone scalars when they
are in the strong coupling regime with fermions.
Only a complete solution of the whole set of Dyson-Schwinger
equations can give us gauge-invariant observables. Thus, the
preceding cross section formula should be read off as only for effective
quantities.
Because of the see-saw suppression, the cross sections mediated
through electroweak
gauge bosons are negligible and, in addition, there is no
$W^{-}W^{+}$ pair production via the $\chi^{0}$ Nambu-Goldstone boson
because $Vertex(\chi^{0}W^{-}W^{+})\equiv 0$ and there is
no Higgs scalar in the theory \cite{Palle1}.
The scaling of the cross section in Eq.(7) is the same as that 
of the cross section $\sigma_{Z}(\bar{N}N\rightarrow \bar{f}f)$
(Eq.(2) of Ref. \cite{Enqvist}): $\sigma_{\chi^{0}} \propto m_{N}^{-2}$.

Knowing the predominant contribution to the total annihilation
cross section, we can estimate the abundances of heavy neutrinos.
The freeze-out temperature of the CDM particle depends on the
cross section only logarithmically \cite{Kolb}, so it is
not neccessary to solve the freeze-out condition \cite{Enqvist}
for $\sigma_{\chi^{0}}$.
We can compare cross sections and make the following
estimates:

\begin{eqnarray*}
s \gg m_{N_{j}}^{2}, m_{\chi^{0}}^{2}, \Gamma_{\chi^{0}}^{2},
\end{eqnarray*}
\begin{eqnarray*}
\Rightarrow \sigma_{\chi^{0}}\simeq \frac{\pi}{\sin ^{4}\theta_{W}}
(\tilde{\alpha_{e}})_{i}(\tilde{\alpha_{e}})_{j}\beta_{N_{i}}
\frac{1}{s},
\end{eqnarray*}
\begin{eqnarray*}
\beta_{N}=(1-4m_{N}^{2}/s)^{1/2},\ s\simeq 4m_{N}^{2}+
6m_{N}T_{f},\ T_{f}\sim \frac{m_{N}}{30},
\end{eqnarray*}
\begin{eqnarray*}
x_{f}\equiv \frac{m_{N}}{T_{f}},\ x_{f}\simeq 17\ 
(m_{N}\simeq 2 GeV),\ x_{f}\simeq 25\ (m_{N}\geq 1 TeV),
\end{eqnarray*}
\begin{eqnarray*}
(\tilde{\alpha_{e}})_{e} \simeq 0.4,\ 
(\tilde{\alpha_{e}})_{\mu} \simeq 1,\ 
(\tilde{\alpha_{e}})_{\tau} \simeq 6,   
\end{eqnarray*}
\begin{eqnarray*}
\Omega_{N} \propto \frac{(n+1)x_{f}^{n+1}}{\sigma_{0}},\ 
\langle\sigma \mid v\mid \rangle\equiv \sigma_{0} x^{-n},
\end{eqnarray*}
\begin{eqnarray*}
\sigma_{Z}(m_{N}\simeq 0.9TeV)\simeq \sigma_{\chi^{0}}(m_{N_{i}}),
\ \Omega_{N} \simeq 1
\end{eqnarray*}
\begin{eqnarray}
\Rightarrow m_{N_{i}}={\cal O}(100 TeV),\ i=\mu ,\tau.
\end{eqnarray}

Although we are limited with the knowledge of the effective 
coupling  of fermions to the Nambu-Goldstone scalars, one
can conclude that heavy Majorana neutrinos with a mass
of order ${\cal O}(100 TeV)$ are good candidates for the CDM
particle.

If one assumes that the radiative decay of the CDM particle dominates,
we can estimate its lifetime from the measurements
of the differential energy flux of the diffuse photon
background \cite{Kribs}:

\begin{eqnarray*}
\Omega_{N} \simeq 1\ and\ Fig.7\ of\ Ref.[9] 
\end{eqnarray*}
\begin{eqnarray}
\Rightarrow 
\tau_{N}={\cal O}(10^{25}s)\ for\ m_{N}={\cal O}(100 TeV).
\end{eqnarray}

The present estimates of the mass and lifetime of the CDM particle
are further challenges for a theory, so the next section 
is devoted to a study of its lifetime.

\section{Lifetime of heavy neutrinos}

Searching for the tree-level decay processes of neutrinos,
one can recall that the tree-level decays of light neutrinos
are kinematically forbidden. Heavy neutrinos couple
to electroweak gauge bosons with see-saw suppression
factors, but in our theory these factors vanish at
tree level \cite{Palle1} because \newline 
 $m_{D_{i}}(tree\ level)\equiv 0$,
i=flavour qauntum number.
Just like for light neutrinos\cite{Moha}, we have to calculate the two-body
radiative flavour changing decay of
heavy neutrinos at the quantum loop order.
 
Let us be concerned with the $N_{i}\rightarrow \nu_{j}\gamma$
decay via the charged weak current loop.
 When dealing with processes containing light and heavy
neutrinos, one has to symmetrize fermion states
with respect to the following interchange of fields:
$\psi_{i} \leftrightarrow \psi^{c}_{i}\Rightarrow f_{i} \leftrightarrow
 F_{i},\ i=flavour$.
Nature should not recognize what we define as a particle or 
an antiparticle. 
 Acknowledging the see-saw 
mixing coefficientes from Eq.(4) and after the particle-antiparticle
symmetrization, one can conclude

\begin{eqnarray*}
\Gamma (N_{i}\rightarrow \nu_{j}\gamma) \propto
(a_{N_{i}}a_{\nu '_{j}}+b_{\nu '_{i}}b_{N_{j}})^{2},\ \forall\ i,j ,
\end{eqnarray*}
\begin{eqnarray}
a_{N_{i}}=-b_{\nu '_{i}}\ and\ a_{\nu '_{j}}=b_{N_{j}} 
\Rightarrow \Gamma (N_{i}\rightarrow \nu_{j}\gamma)\equiv 0.
\end{eqnarray}
 
We shall now study the decay processes $N_{i}\rightarrow
N_{j}\gamma$ induced by the loop exchange of the W weak boson and
charged leptons. The effective transition operator will be 
estimated preturbatively, so one has to choose the
renormalization conditions for the flavour mixing of leptons.

The usual wisdom for fermion mixing is to apply the on-shell
fermion mixing scheme of Ref.\cite{Aoki}. It has been shown that 
in flavour-changing lepton radiative processes, this scheme
causes the appearance of the dimension-four operators \cite{Palle2}
besides the standard dimension five-operators \cite{Moha}.
However, the authors of Ref.\cite{Gambino} have recently shown
that the on-shell renormalization scheme is not consistent
because it violates Ward-Takahashi identities and leads to
gauge-dependent physical amplitudes. Instead, one can introduce
a natural and consistent
 prescription where flavour-changing
self-energies vanish at zero momentum\cite{Gambino}. 
From the Ward-Takahashi identity and the conservation
of the electromagnetic current one can conclude that
new renormalization conditions of Ref.\cite{Gambino}
 do not induce the appearance of four-dimensional
operators in flavour changing lepton radiative decays.

The effective see-saw suppressed transition operator 
for $N_{i}\rightarrow N_{j}\gamma,\ (i,j=flavours)$ is
evaluated in the t'Hooft-Feynman gauge (loops with
exchanged $WWl$ and $Wll$, l=charged lepton) because
only a strongly coupled heavy neutrino-Nambu-Goldstone scalar 
(decoupled in the unitary gauge) system can produce
the mass splitting of heavy neutrinos
(CP invariance assumed, \cite{Moha}):

\begin{eqnarray}
{\cal A}_{\mu}=[H(0)+H_{5}(0)\gamma_{5}]\sigma_{\mu\rho}(p_{2})^{\rho},
\end{eqnarray}
\begin{eqnarray*}
opposite\ CP\ eigenvalues\ of\ neutrinos:
\end{eqnarray*}
\begin{eqnarray*}
H=\frac{eg^{2}}{16\pi ^{2}}\sqrt{\frac{m_{\nu_{i}}m_{\nu_{j}}}
{m_{N_{i}}m_{N_{j}}}}
\sum_{l}U^{*}_{lj}U_{li}
(H_{L}+H_{R}), H_{5}=0, 
\end{eqnarray*}
\begin{eqnarray*}
same\ CP\ eigenvalues\ of\ neutrinos:
\end{eqnarray*}
\begin{eqnarray*} 
H_{5}&=&\frac{eg^{2}}{16\pi ^{2}}\sqrt{\frac{m_{\nu_{i}}m_{\nu_{j}}}
{m_{N_{i}}m_{N_{j}}}}
\sum_{l}U^{*}_{lj}U_{li}
(H_{R}-H_{L}), H=0,
\end{eqnarray*}
\begin{eqnarray*}
H_{L}&=&m_{N_{j}}(C_{0}+C_{11}+C_{12}+C_{23}), \\
H_{R}&=&m_{N_{i}}(-C_{0}+C_{12}+\tilde{C}_{12}
-2C_{11}-\frac{1}{2}\tilde{C}_{11}-C_{21}+C_{23}), \\
C_{0}&=&C_{0}(p_{1}^{2},p_{2}^{2},p_{3}^{2};M_{W},m_{l},m_{l}), \\
\tilde{C}_{0}&=&C_{0}(p_{1}^{2},p_{2}^{2},p_{3}^{2};m_{l},
M_{W},M_{W}), \\
p_{1}^{2}&=&m_{N_{i}}^{2},\ p_{2}^{2}=0,\ p_{3}^{2}=m_{N_{j}}^{2}, 
m_{l}=charged\ lepton\ mass, \\
&&for\ further\ definitions\ see\ Appendix\ A.
\end{eqnarray*}

The amplitude is ultraviolet(UV) finite, but it contains 
infrared(IR) singularity in the limes $m_{l}\rightarrow 0$.
It can be visualized that the IR singularity comes from
the following two Green's functions:

\begin{eqnarray*}
\Re B_{0}(0;m_{l},m_{l})&=&-\ln m_{l}^{2}+... ,\hspace*{50 mm}\\
\Re C_{0}&=&\frac{1}{p_{1}^{2}-p_{3}^{2}}(
\frac{1}{2}\ln ^{2}w_{1}-\frac{1}{2}\ln ^{2}w_{3})+..., \\
w(p^{2},M_{W},m_{l})&=&\frac{1}{2p^{2}}
(p^{2}-M_{W}^{2}+m_{l}^{2}-\sqrt{(p^{2}-M_{W}^{2}
+m_{l}^{2})^{2}-4p^{2}m_{l}^{2}}), \\
w_{1}&=&w(p_{1}^{2},M_{W},m_{l}),\ w_{3}=w(p_{3}^{2},M_{W},m_{l}).
\end{eqnarray*}

The IR singularity can be removed in a natural way through
the leptonic GIM mechanism:

\begin{eqnarray*}
H=\lim_{M\rightarrow 0}\sum_{l}[{\cal H}(m_{l}^{2},...)
-{\cal H}(m_{l}^{2}=M^{2},...)], \\
similarly\ for\ H_{5}.
\end{eqnarray*}

The IR singular and constant terms of the whole amplitude are
now subtracted away.

Instead of the Higgs mechanism, the finite
UV scale is introduced into the theory, so one has to study
the finite-scale effects. They enter into the calculations 
through the evaluations of scalar Green's functions \cite{Palle3}.
One can naively expect large corrections, but the explicit
evaluation (see Appendix B) tells us that they are three
orders of magnitude smaller,
 thus much smaller than uncertainties
in masses. We shall neglect this effect in the numerical
estimates of lifetimes.
 The reason for small corrections is that for these decay 
processes we need the knowledge of
the Green's functions in the timelike region for
very high masses ($m_{N_{i}}\gg \Lambda$).
The contribution from the timelike region dominates over
that of the spacelike region.
On the contrary, in QCD
one studies the spacelike region with a cutoff $\Lambda$,
thus a large deviation will be encountered \cite{Palle3}
for the scale $\mu \geq \Lambda$.

A straightforward calculation gives the partial decay width
of the Majorana heavy neutrino(assumed the same CP eigenvalues of neutrinos
\cite{Moha}):

\begin{eqnarray}
\Gamma(N_{i}\rightarrow N_{j}\gamma)=
\frac{(m_{N_{i}}^{2}-m_{N_{j}}^{2})^{3}}{8\pi m_{N_{i}}^{3}}
\mid H_{5}\mid ^{2}.
\end{eqnarray}

Because of the fact that $m_{N_{i}}^{2} \gg M_{W}^{2}$, 
a partial decay width
is insensitive to the weak boson mass to the leading order:

\begin{eqnarray}
\Gamma(N_{i}\rightarrow N_{j}\gamma) \propto m_{N_{i}}
\sin ^{2}(2\theta_{ij})(\frac{m_{\nu_{i}}m_{\nu_{j}}}
{m_{N_{i}}m_{N_{j}}})(\frac{m_{l_{i}}}{m_{N_{j}}})^{4}.
\end{eqnarray}

Owing to this scaling,
the gauge dependence of the effective transition operator is 
of subleading order and there is no need to study 
the gauge cancellations in detail.  The
heavy-quark symmetry of Isgur and Wise or the Appelquist-Carazzone
decoupling theorem are also examples in field theory where
a good approximation in a certain sector of the complete
theory describes the actual
physical situation correctly.

Next section is devoted to final numerical evaluations and physical
conclusions.

\section{Results and conclusions}

From the requirement for the abundance of the CDM particle
and the requirement of the cosmic diffuse photon
background we extracted the information about possible
masses and lifetimes of heavy neutrinos.
The general scaling of the lifetime on light- and
heavy-neutrino massees and mixing angles suggests that there is 
no dependence on the mass of the decaying particle, because
of the cancellation(see Eq.(13)).

Now we present our numerical results with an appropriate
choice of neutrino masses and mixing angles. From
the scaling relation Eq.(13) one can easily estimate
a partial decay width for some different set of parameters.

\begin{eqnarray*}
m_{N_{e}}=10 TeV,\ m_{N_{\mu}}=50 TeV,\ m_{N_{\tau}}=200 TeV,
\end{eqnarray*}
\begin{eqnarray*}
m_{\nu_{e}}=0.03 eV,\ m_{\nu_{\mu}}=0.3 eV,\ 
m_{\nu_{\tau}}=2 eV,
\end{eqnarray*}
\begin{eqnarray*}
\theta_{12}=0.21,\ \theta_{23}=0.02,\ \theta_{31}=0.002,
\end{eqnarray*}
\begin{eqnarray*}
\tau=[\sum_{a}\tau^{-1}_{a}]^{-1},
\end{eqnarray*}
\begin{eqnarray*}
\tau (N_{\tau}\rightarrow N_{\mu}\gamma)=2.3\cdot 10^{25}s,\ 
\tau (N_{\tau}\rightarrow N_{e}\gamma)=5.4\cdot 10^{24}s,
\end{eqnarray*}
\begin{eqnarray}
\tau (N_{\mu}\rightarrow N_{e}\gamma)=2.0\cdot 10^{26}s.
\end{eqnarray}

The lack of our knowledge of the mixing angles and neutrino 
masses does not prevent us from drawing main conclusions of this paper
concerning heavy neutrinos: \newline
(1) Acknowledging the see-saw relation for light-neutrino
masses and the relation for the mixing angles $\theta_{W}=
2 (\theta_{12}+\theta_{23}+\theta_{31})$ \cite{Palle1},
the  (BY) theory predicts that the heavy neutrino $N_{\mu}$ 
or $N_{\tau}$ with the
mass ${\cal O}(100 TeV)$ and the lifetime $\tau={\cal O}
(10^{25})s$ is a CDM particle which can solve the problem
of the cosmic diffuse photon background; this means that a
further detailed astronomical study of the diffuse photon background
could be useful for the study of the dynamics of the CDM particle,
which could be supplemented by terrestrial measurements
(for example, the DAMA experiment at the Gran Sasso National Laboratory);
it is not excluded that processes with heavy neutrinos could
explain the recently observed(EGRET) diffuse gamma halo around
our Galaxy. \newline
(2) Essential ingredients of the (BY) theory, which gives 
successful phenomenological answers, are
the peculiar
chiral structure of the theory, the absence of the Higgs
scalar and the presence of the finite UV scale to 
fix the scale of heavy neutrinos and weak gauge bosons,
as well as the presence of the UV finite self-consistent bootstrap
system of equations leading to the finite number of fermion
families.
 \newline
 (3) Because of the rather strong tree-level coupling of
 $N_{i}$ to $\chi^{0}$: $\frac{\alpha_{e}}{4\sin ^{2}      
\theta_{W}}(\frac{m_{0}}{M_{W}})^{2}=0.29,$
$m_{0}=485 GeV$, it seems that even the lightest one
of heavy neutrinos acquires the mass mostly from
the quantum loop ($m_{N_{e}} \gg m_{0}$).
 \newline
 (4) The forthcoming higher-luminosity measurements at Tevatron
could
  confirm the nonresonant enhancement of the
QCD amplitudes beginning in the vicinity of $\Lambda$ 
($\lim_{\mu \rightarrow \infty} \alpha_{s}^{\Lambda}(\mu)\neq 0$)
\cite{Palle3}. Also note the unsettled problem of the
anomalous b-quark weak coupling (LEP and SLD) as a possible
consequence of the finite scale effect in the loop corrections
with the t-quark exchange \cite{Palle3}. \newline
(5) The Einstein-Cartan cosmology requires that $\Omega_{m}\simeq 2$,
 thus the CDM particle must be the spinning particle (fulfilled
for the (BY) theory) \cite{Palle4}; this setting of the average
mass density of the Universe is a consequence of the following bootstrap
at the level of the Einstein-Cartan equations: the number
density of the particle that dominates the energy momentum density,
at the same time dominates the spin density of the Universe, but 
the energy momentum and spin of matter are coupled
to the curvature and torsion of spacetime with the same
coupling constant. \newline
 (6) The current controversies \cite{Silk} concerning the 
average mass density of
the Universe could be resolved only with the most general
cosmological model that contains cold and hot dark matter,
cosmological constant, 
baryons and CMBR and the metric not only with expansion, but also
with a small amount of vorticity, acceleration and shear \cite{Palle4};
namely, one can solve the primordial mass density fluctuation
within the Einstein-Cartan cosmology only beyond the standard
Friedmann-Lema$\hat{i}$tre-Robertson-Walker metric \cite{Palle5};
it was shown a long time ago that the primordial vorticity could generate 
a cosmic magnetic field \cite{Harrison}.
\newline
\hspace*{59mm}$\ *\ *\ *\ $
\newline
This work was supported by the Ministry of Science and Technology of
the Republic of Croatia under Contract No. 00980103.
\newline
\section*{Acknowledgement}
I should like to thank Drs. Kribs and Rothstein for calling
my attention to their article.

\section*{Appendix A}

Here we display the scalar and tensor Green's functions used 
in the evaluation of the effective transition flavour-changing 
 operator: 

\begin{eqnarray*}
A(m)=\frac{1}{\imath\pi^{2}}\int d^{4}q
\frac{1}{q^{2}-m^{2}+\imath\varepsilon},   \\
\end{eqnarray*}
\begin{eqnarray*}
B_{0}(p^{2};m_{1},m_{2})=\frac{1}{\imath\pi^{2}}\int d^{4}q
\frac{1}{(q^{2}-m_{1}^{2}+\imath\varepsilon)
((q+p)^{2}-m_{2}^{2}+\imath\varepsilon)},  \\
\end{eqnarray*}
\begin{eqnarray*}
p_{\mu}B_{1}(p^{2};m_{1},m_{2})=\frac{1}{\imath\pi^{2}}
\int d^{4}q\frac{q_{\mu}}{(q^{2}-m_{1}^{2}+
\imath\varepsilon)((q+p)^{2}-m_{2}^{2}+\imath\varepsilon)},  \\
\end{eqnarray*}
\begin{eqnarray*}
g_{\mu\nu}B_{22}+p_{\mu}p_{\nu}B_{21}=
\frac{1}{\imath\pi^{2}}
\int d^{4}q\frac{q_{\mu}q_{\nu}}{(q^{2}-m_{1}^{2}+
\imath\varepsilon)((q+p)^{2}-m_{2}^{2}+\imath\varepsilon)}, \\
\end{eqnarray*}
\begin{eqnarray*}
p_{1}+p_{2}+p_{3} = 0,
\end{eqnarray*}
\begin{eqnarray*}
C_{0}(p_{1}^{2},p_{2}^{2},p_{3}^{2};m_{1},m_{2},m_{3})=
\frac{1}{\imath\pi^{2}}\int 
\frac{d^{4}q}{(q^{2}-m_{1}^{2}+\imath\varepsilon)((q+p_{1})^{2}-m_{2}^{2}+
\imath\varepsilon)((q-p_{3})^{2}-m_{3}^{2}+\imath\varepsilon)}, \\
\end{eqnarray*}
\begin{eqnarray*}
p_{1}^{\mu}C_{11}+p_{2}^{\mu}C_{12}=
\frac{1}{\imath\pi^{2}}\int
d^{4}q\frac{q^{\mu}}{(q^{2}-m_{1}^{2}+\imath\varepsilon)
((q+p_{1})^{2}-m_{2}^{2}+\imath\varepsilon)
((q-p_{3})^{2}+m_{3}^{2}+\imath\varepsilon)}, \\
\end{eqnarray*}
\begin{eqnarray*}
g^{\mu\nu}C_{24}+p_{1}^{\mu}p_{1}^{\nu}C_{21}+p_{2}^{\mu}
p_{2}^{\nu}C_{22}+(p_{1}^{\mu}p_{2}^{\nu}+p_{2}^{\mu}
p_{1}^{\nu})C_{23} \\
=\frac{1}{\imath\pi^{2}}\int
d^{4}q\frac{q^{\mu}q^{\nu}}{(q^{2}-m_{1}^{2}+\imath\varepsilon)
((q+p_{1})^{2}-m_{2}^{2}+\imath\varepsilon)
((q-p_{3})^{2}+m_{3}^{2}+\imath\varepsilon)}, \\
\end{eqnarray*}
\begin{eqnarray*}
B_{1}(p^{2};m_{1},m_{2})=\frac{1}{2 p^{2}}[A(m_{1})-A(m_{2})+
(m_{2}^{2}-m_{1}^{2}-p^{2})B_{0}(p^{2};m_{1},m_{2})], \\
B_{22}(p^{2};m_{1},m_{2})=\frac{1}{6}[A(m_{2})+
2 m_{1}^{2}B_{0}(p^{2};m_{1},m_{2})+
(p^{2}+m_{1}^{2}-m_{2}^{2})B_{1}(p^{2};m_{1},m_{2})], \\
B_{21}(p^{2};m_{1},m_{2})=\frac{1}{3 p^{2}}[A(m_{2})
-m_{1}^{2}B_{0}(p^{2};m_{1},m_{2})
-2(p^{2}+m_{1}^{2}-m_{2}^{2})B_{1}(p^{2};m_{1},m_{2})],
\end{eqnarray*}
\begin{eqnarray*}
C_{12}=\frac{1}{(p_{1}\cdot p_{2})^{2}-p_{1}^{2}p_{2}^{2}}
[p_{1}^{2}\Sigma_{2}+(p_{1}^{2}+p_{1}\cdot p_{2})\Sigma_{1}], 
\end{eqnarray*}
\begin{eqnarray*}
C_{11}=\frac{1}{p_{1}^{2}}(\Sigma_{1}-p_{1}\cdot p_{2}C_{12}), 
\end{eqnarray*}
\begin{eqnarray*}
\Sigma_{1}=\frac{1}{2}B_{0}(p_{3}^{2};m_{1},m_{3})+
\frac{1}{2}(-p_{1}^{2}-m_{1}^{2}+m_{2}^{2})C_{0}(p_{1}^{2},
p_{2}^{2},p_{3}^{2};m_{1},m_{2},m_{3})-
\frac{1}{2}B_{0}(p_{2}^{2};m_{2},m_{3}), \\
\Sigma_{2}=-\frac{1}{2}B_{0}(p_{1}^{2};m_{1},m_{2})+
\frac{1}{2}(p_{3}^{2}-m_{3}^{2}+m_{1}^{2})C_{0}(p_{1}^{2},
p_{2}^{2},p_{3}^{2};m_{1},m_{2},m_{3})+\frac{1}{2}
B_{0}(p_{2}^{2};m_{2},m_{3}),
\end{eqnarray*}
\begin{eqnarray*}
C_{23}=\frac{1}{2(p_{1}^{2}p_{2}^{2}-(p_{1}\cdot p_{2})^{2})}
[(-3p_{1}^{2}-4p_{1}\cdot p_{2})\Lambda_{2}-3p_{1}^{2}\Lambda_{4}
+p_{1}\cdot p_{2}\Lambda_{1}-p_{2}^{2}\Lambda_{3}], 
\end{eqnarray*}
\begin{eqnarray*}
C_{21}=-\frac{1}{p_{1}\cdot p_{2}}[\Lambda_{2}+
\Lambda_{4}+p_{2}^{2}C_{23}], 
\end{eqnarray*}
\begin{eqnarray*}
C_{22}=\frac{1}{p_{1}\cdot p_{2}}[\Lambda_{3}-
p_{1}^{2}C_{23}], 
\end{eqnarray*}
\begin{eqnarray*}
C_{24}=\Lambda_{2}-p_{1}^{2}C_{21}-p_{1}\cdot p_{2}C_{23}, 
\end{eqnarray*}
\begin{eqnarray*}
\Lambda_{1}=B_{0}(p_{2}^{2};m_{2},m_{3})+m_{1}^{2}C_{0}, 
\end{eqnarray*}
\begin{eqnarray*}
\Lambda_{2}=\frac{1}{2}B_{1}(p_{3}^{2};m_{1},m_{3})+
\frac{1}{2}(-p_{1}^{2}+m_{2}^{2}-m_{1}^{2})C_{11}, 
\end{eqnarray*}
\begin{eqnarray*}
\Lambda_{3}=\frac{1}{2}B_{1}(p_{3}^{2};m_{1},m_{3})-
\frac{1}{2}B_{1}(p_{2}^{2};m_{2},m_{3})+
\frac{1}{2}(-p_{1}^{2}+m_{2}^{2}-m_{1}^{2})C_{12}, 
\end{eqnarray*}
\begin{eqnarray*}
\Lambda_{4}=-\frac{1}{2}B_{1}(p_{1}^{2};m_{1},m_{2})+
\frac{1}{2}(p_{3}^{2}-m_{3}^{2}+m_{1}^{2})C_{11}.
\end{eqnarray*}

\section*{Appendix B}

The real parts of the two- and three-point scalar Green's 
functions in the noncontractible space are given
as in Ref. \cite{Palle3}:

\begin{eqnarray*}
\Re B_{0}^{\Lambda}(p^{2};m_{1},m_{2})=
(\int_{0}^{\Lambda^{2}}d y K(p^{2},y)+\theta (p^{2}-m_{2}^{2}) 
 \int_{-(\sqrt{p^{2}}-m_{2})^{2}}
^{0}d y \Delta K(p^{2},y) )\frac{1}{y+m_{1}^{2}}, \\
K(p^{2},y)=\frac{2 y}{-p^{2}+y+m_{2}^{2}+
\sqrt{(-p^{2}+y+m_{2}^{2})^{2}+4 p^{2} y}},  \hspace*{40 mm}\\
\Delta K(p^{2},y)=\frac{\sqrt{(-p^{2}+y+m_{2}^{2})^{2}+4 p^{2} y}}{p^{2}}.
 \hspace*{60 mm}
\end{eqnarray*}

The integration in the second term is performed from the branch 
point of the square root $\sqrt{(-p^{2}+y+m_{2}^{2})^{2}+4 p^{2} y}\equiv 
\imath Z$ and the additional kernel is derived as the difference:
$ \Delta K(p^{2},y)=K(p^{2},y)-K^{*}(p^{2},y)=\frac{2 y}{-p^{2}+y+
m_{2}^{2}+\imath Z}-\frac{2 y}{-p^{2}+y+m_{2}^{2}-\imath Z}$.

The integration over singularities is supposed to be the principal-value 
integration.

In the case of the two-point Green's function $B_{0}^{\Lambda}$,
 we need the explicit form of the additional term for the integration
in the timelike region because the integration 
in the spacelike region is divergent in the limes $\Lambda 
 \rightarrow \infty$. However, the three-point scalar Green's
functions are UV-convergent and we do not need to know the explicit
form of the additional terms because they do not depend on the 
UV cut-off and we can use the analytical continuation of the standard
 Green's functions written in terms of the dilogarithms\cite{Palle3,skalar}:

\begin{eqnarray*}
\Re C_{0}^{\Lambda}(p_{i},m_{j})=\int_{0}^{\Lambda^{2}}dq^{2} \Phi 
(q^{2},p_{i},m_{j})
+\int_{TD}dq^{2} \Xi (q^{2},p_{i},m_{j}), \hspace*{10 mm}\\
\Re C_{0}^{\Lambda}(p_{i},m_{j})=Re C_{0}^{\infty}(p_{i},m_{j})-
\int^{\infty}_{\Lambda^{2}}d q^{2} \Phi (q^{2},p_{i},m_{j}), 
 \hspace*{20 mm} \\
\Phi \equiv function\ derived\ by\ the\ angular\ integration\ 
after\ Wick's\ rotation,\\
C_{0}^{\infty}\equiv standard\ 't\ Hooft-Veltman\ scalar\ function, 
\hspace*{20 mm} \\
TD\equiv timelike\ domain\ of\ integration. \hspace*{40 mm}
\end{eqnarray*}

This equation is valid for arbitrary external momenta. The same formula
 is applicable to the higher n-point one loop scalar Green's functions.
 
To confirm the claim that the finite scale effects in the
heavy-neutrino radiative decay are small, we evalute 
 the $C \equiv \tilde{C}_{0}$ Green's function as an example:

\begin{eqnarray*}
\Re C^{\Lambda}=\Re C^{\infty} - \bigtriangleup \Gamma,
\end{eqnarray*}
\begin{eqnarray*}
\bigtriangleup \Gamma = \Gamma^{\infty}-\Gamma^{\Lambda},
\end{eqnarray*}
\begin{eqnarray*}
\Gamma^{\Lambda}=-\frac{2}{\pi}\int^{\Lambda}_{0} d q
\frac{q^{3}}{q^{2}+M_{W}^{2}}(I_{1}+I_{2}), 
\end{eqnarray*}
\begin{eqnarray*}
I_{1}=\int^{+1}_{-1} d x \frac{1}{(q^{2}+m_{1}^{2}+m_{l}^{2})^{2}
+4m_{1}^{2}q^{2}x^{2}}\frac{-m_{1}\mid x\mid}{k} 
\end{eqnarray*}
\begin{eqnarray*}
\times (\arctan\frac{q^{2}+m_{2}^{2}+m_{l}^{2}+2kq\sqrt{1-x^{2}}}
{2(m_{1}+k)q\mid x\mid}-
\arctan\frac{q^{2}+m_{2}^{2}+m_{l}^{2}-2kq\sqrt{1-x^{2}}}
{2(m_{1}+k)q\mid x\mid}), 
\end{eqnarray*}
\begin{eqnarray*}
I_{2}=\int^{+1}_{-1} d x \frac{1}{(q^{2}+m_{1}^{2}+m_{l}^{2})^{2}
+4m_{1}^{2}q^{2}x^{2}}\frac{q^{2}+m_{1}^{2}+m_{l}^{2}}{4kq} 
\end{eqnarray*}
\begin{eqnarray*}
\times \ln\frac{(q^{2}+m_{2}^{2}+m_{l}^{2}+2kq\sqrt{1-x^{2}})^2
+4(m_{1}+k)^{2}q^{2}x^{2}}{(q^{2}+m_{2}^{2}+m_{l}^{2}
-2kq\sqrt{1-x^{2}})^{2}+4(m_{1}+k)^{2}q^{2}x^{2}} ,
\end{eqnarray*}
\begin{eqnarray*}
k\equiv \frac{m_{2}^{2}-m_{1}^{2}}{2m_{1}},
\end{eqnarray*}
\begin{eqnarray*}
p_{1}^{2}=m_{1}^{2}=(10 TeV)^{2},\ p_{3}^{2}=m_{2}^{2}=(0.5 TeV)^{2},\ 
\Lambda = 326 GeV,
\end{eqnarray*}
\begin{eqnarray*}
\bigtriangleup \Re C^{\infty}\equiv \Re C^{\infty}(m_{l}=
m_{\mu})-\Re C^{\infty}(m_{l}=m_{e})=1.00\cdot 10^{-8}TeV^{-2},
\end{eqnarray*}
\begin{eqnarray*}
\bigtriangleup (\bigtriangleup \Gamma)\equiv 
\bigtriangleup \Gamma(m_{l}=m_{\mu})-
\bigtriangleup \Gamma(m_{l}=m_{e})=-6.87\cdot 10^{-12}TeV^{-2}
\end{eqnarray*}
\begin{eqnarray*}
\Rightarrow \mid \bigtriangleup\Re C^{\infty}\mid \gg \mid \bigtriangleup 
(\bigtriangleup \Gamma)\mid.
\end{eqnarray*}

\end{document}